\documentclass[english,prd,twocolumn,superscriptaddress,nofootinbib,preprintnumbers]{revtex4-1}
\usepackage[latin1]{inputenc}
\usepackage{graphicx}
\usepackage{amssymb}
\usepackage{color}
\usepackage{float}
\usepackage{amsmath}
\usepackage{amsfonts}
\usepackage{dcolumn}
\usepackage{hyperref}
\usepackage{subfigure}

\usepackage{mathtools}
\usepackage{soul,xcolor}
\hypersetup{
     breaklinks=true,
    pdfstartview={FitH},    
    colorlinks=true,       
    linkcolor=blue,          
    citecolor=red,        
    filecolor=magenta,      
    urlcolor=blue,           
    anchorcolor=green,      
    linktocpage=true
}

\newcommand{\be}{\begin{equation}}
\newcommand{\ee}{\end{equation}}
\newcommand{\bea}{\begin{eqnarray}}
\newcommand{\eea}{\end{eqnarray}}

\def\G4{\bar{G_4}}
\newcommand*\DAlambert{\mathop{}\!\mathbin\Box}
\newcommand{\Mpl}{M_{\textrm{Pl}}}
\renewcommand{\(}{\left(}
\renewcommand{\)}{\right)}
\renewcommand{\k}{\vec{k}}

\newcommand{\nn}{\nonumber}
\def\al{\alpha}
\def\bet{\beta}
\def\gam{\gamma}

\def\Om{\Omega}
\def\sig{\sigma}

\def\lam{\lambda}
\def\ep{\epsilon}

\def\S{\mathcal{S}}

\def\doi{http://doi.org}
\def\arxiv{http://arxiv.org/abs}

\def\r{\mathrm{r}}
\def\g{\mathrm{g}}
\def\f{\mathrm{f}}

\def\m{\mathrm{m}}

\def\s{\mathrm{s}}
\def\d{\mathrm{d}}

\def\k{\kappa_\phi}

\allowdisplaybreaks

\begin{document}

\title{Bigravity and Horndeski gravity connected by a disformal coupling}

\author{Radouane Gannouji}
\email{radouane.gannouji@pucv.cl}
\affiliation{Instituto de F\'{\i}sica, Pontificia Universidad  Cat\'olica de Valpara\'{\i}so, Casilla 4950, Valpara\'{\i}so, Chile}
\author{Md. Wali Hossain}
\email{wali.hossain@apctp.org}
\affiliation{Asia Pacific Center for Theoretical Physics, Pohang 37673, Korea}
\author{Nur Jaman}
\email{nurjaman@ctp-jamia.res.in}
\affiliation{Centre for Theoretical Physics, Jamia Millia Islamia, New Delhi, India}
\author{M. Sami}
\email{samijamia@gmail.com}
\affiliation{Centre for Theoretical Physics, Jamia Millia Islamia, New Delhi, India}
\affiliation{Maulana Azad National Urdu
University, Gachibowli, Hyderabad-500032,India}\affiliation{Institute for Advanced Physics $\&$ Mathematics, Zhejiang University of Technology, \\Hangzhou, 310032, China}

\begin{abstract}
In this paper, we have studied bi-gravity theory in a very specific limit where we focused  only on one degree of freedom generated by the massive graviton. We have analyzed the model in the context of cosmology and demonstrated that the model can give rise to late time cosmic acceleration as an attractor of the dynamical system. However, the observational constraints due to tensor perturbations are stringent giving rise to  large fine tuning.  
\end{abstract}

\maketitle

\section{Introduction}
Since the first days of General Relativity (GR), alternatives to the standard model in gravity have been proposed; 
the last century has been very prolific and creative in model building. Some of the models have been introduced by fundamental considerations such as the renormalization of GR or the hierarchy problem whereas the others have been motivated by phenomenological arguments from cosmology such as origin of the accelerated expansion of the universe. Far from purely theoretical challenge to construct a consistent theory beyond GR, these frameworks have been essential for the check of consistency  of all assumptions behind the theory of GR. For example, alternative theories such as Brans-Dicke\cite{Brans:1961sx} model played a primary role in the elaboration of new experiments which ironically led to the demise of such theories. With the same ambition, modern alternatives to the standard model challenge GR and therefore all the fundamental assumptions behind Einstein's theory. In fact, according to the Lovelock theorem\cite{Lovelock:1972vz}, which assumes 4 postulates , the theory of GR is a unique consistent metric theory of gravity. Any modification of gravity challenges therefore one or many of these assumptions which are 1- the theory is defined by a unique field, the metric, 2- the theory is defined in 4 dimensions, 3- the theory is defined by a second order differential equation, 4- the theory is invariant under diffeomorphism. Massive gravity, (for  a review see\cite{massive}) as an alternative to GR, breaks one of the fundamental ideas behind this uniqueness theorem, the invariance under the symmetry group by considering a mass term to one of the fundamental particles in physics; the graviton. This mass could arise from a Higgs mechanism even though a consistent implementation of this process is unknown. So far, the mass has been added by hand. At the linear order, the theory developed by Fierz and Pauli in 1939\cite{Fierz:1939ix}
 \,\ is well defined and doesn't suffer from fundamental problems until we couple the theory to matter. In fact, a massive spin-2 field carries 5 degrees of freedom such that the coupling of the longitudinal mode is not suppressed  even in the decoupling limit and gives a theory different from GR. The solution to this problem was provided by Vainshtein\cite{Vainshtein:1972sx} by including non-linear effects that wakes up the sixth degree which is a ghost. This cumbersome construction found finally few solutions by either considering a model of soft massive gravity ({\it \`a la} DGP)\cite{Dvali:2000hr} where the graviton has no single massive pole or by considering a very specific ``potential'' to the theory known as dRGT massive gravity\cite{deRham:2010kj,deRham:2010ik}. Unfortunately, this ghost free theory of massive gravity failed to provide a standard cosmology \cite{DeFelice:2012mx,deRham:2010ik}. Also it is not clear why the theory should be defined over a fixed Minkowsky spacetime. Following this development, first some new background metrics were proposed such as de Sitter space time \cite{deRham:2012kf} and therefore implicitly pointing out the problem in the freedom to choose a fixed reference space time. In this context, bi-gravity where the reference metric acquires a dynamics became a new theory to be explore. This new framework introduced in \cite{Hassan:2011zd} became rapidly very popular but if one assumes it as a good candidate of dark energy(for a review see\cite{Darkenergy}), one requires the mass term in the theory to be of the order of the Hubble rate today $(H_0)$. However this would imply a gradient instability during radiation era \cite{Comelli:2012db}. Therefore making us forced to take a large mass of the graviton ($m\gg H_0$) which either would make the theory indistinguishable from GR at low energy or inconsistent with local tests because the Vainshtein mechanism would be insufficient to suppress the modification. In this context, it was proposed \cite{DeFelice:2017oym} to add to the Vainshtein effect, a chameleon mechanism by promoting the mass term to a function that allows for the mass of the massive graviton to be environment dependent.
 
In this framework, one always deals with a particular degree of freedom, a scalar field which couples to matter and provides strong information of the viability of the theory, for instance, in the decoupling limit or the construction of Galileon theory by the brane bending mode. In this context, we propose to study a particular case, which is a sub-category of the extended massive gravity proposed in \cite{DeFelice:2017oym} where the two metrics are related to each other by a disformal transformation\footnote{In ref\cite{DeFelice:2017oym}, the environment dependency is implemented by promoting $\beta_i$ as a function of scalar field $\phi$ and considering an additional kinetic term of $\phi$. We have done the same but instead of considering additional kinetic term, we imagined that it is one of the modes of the massive graviton that we have isolated from other modes by this disformal coupling.}. We will show that the theory will reduce to the quartic DBI Galileon which has been first introduced through extra dimensions \cite{deRham:2010eu} and therefore connecting brane models with the presence of Gauss-Bonnet to bi-gravity in 4D in a ``disformal limit''. Finally we will study the cosmology of the model.

The structure of this paper is as follows. In Sec.~\ref{theAction} we describe the scenario under consideration. In Sec.~\ref{Cosmologysec} we study the cosmological dynamics and checked the stability of the system at the fixed points through dynamical system analysis. In  Sec.~\ref{TenPert} we study the tensor perturbations and compared the model with recent observational bounds on the gravitational wave speed. In Sec.~\ref{Chame} we modify our system to a general set-up of $\phi$ dependent $\beta$ with addition of canonical kinetic term for it and briefly discuss the its cosmology. Finally we conclude in Sec.~\ref{concl}.

\section{The Action}
\label{theAction}
We consider the following bimetric theory \cite{Hassan:2011zd}
\begin{align}
\S=&\frac{M_g^2}{2}\int \d^4x \sqrt{-g} R[g] +\epsilon\frac{M_f^2}{2}\int \d^4x \sqrt{-f} R[f]  \nonumber\\
&+ \frac{M_g^2}{2}m^2 \int \d^4x \sqrt{-g}\sum_{i=0}^{4}\beta_i U_i[\s] \, ,
\label{eq:action0}
\end{align}
where $s^\alpha _\beta \equiv(\sqrt{g^{-1}f})^\alpha _{\beta}$ and the potentials $U_i(i=0,1,2,3,4)$ are defined in terms of $T_n \equiv Tr[s^n]$ as 
\begin{align}
 U_0&=1\, , \\
 U_1&= T_1\, , \\
 U_2&= \frac{1}{2}(T_1^2-T_2)\, , \\
 U_3&= \frac{1}{6}(T_1^3-3T_2T_1+2T_3) \, , \\
 U_4&= \frac{1}{24}(T_1^4-6T_1^2T_2+3T_2^2+8T_1T_3-6T_4) \, .
\end{align}
where we have added in the action $\epsilon=\pm 1$ for completeness. 

In this paper we consider the disformal relation between the two metrics
\begin{eqnarray}
f_{\mu\nu}= g_{\mu\nu}+B(\phi)\phi_{\mu}\phi_{\nu} \, ,
\label{eq:dis}
\end{eqnarray}
where $\phi_{\mu}\equiv\partial_{\mu}\phi=\nabla_\mu \phi$. From which we can find easily the relations

\begin{align}
s^\alpha _\beta s^\beta _\gamma&= \delta^\alpha _\gamma +B(\phi)\phi^\alpha \phi_{\gamma} \, ,
\label{eq:s_albet}\\
f^{\mu\nu}&=g^{\mu\nu}+\frac{B}{2BX-1}\phi^\mu\phi^\nu\\
\Gamma^\rho _{\mu\nu}[f] &= \Gamma^{\rho}_{\mu\nu}[g]-\frac{B'}{2(2BX-1)}\phi^\rho \phi_{\mu}\phi_{\nu}-\frac{B}{2BX-1}\phi_{\mu\nu}\phi^\rho\\
R[f]&= R[g]-\frac{B}{1-2BX}R_{\mu\nu}\phi^\mu\phi^\nu+\nabla_\mu V^\mu, 
\label{eq:Rf}
\end{align}
where 
\begin{align}
V^\mu=\frac{B}{1-2BX}\Big[\DAlambert\phi \phi^\mu-\phi^{\mu\nu}\phi_{\nu}\Big],
\end{align}
where $\phi_{\al\bet}\equiv\nabla_\bet\nabla_\al\phi$, $X=-\frac{1}{2}\phi^{\mu}\phi_\mu$ and we use the metric $g_{\mu \nu}$ to lower and upper the indices. Here we can say that $\beta_i$ can be regarded as function of the same  scalar field as $\phi$ in the disformal relation~(\ref{eq:dis}), without loss of generality. From Eqs.~(\ref{eq:s_albet}) and (\ref{eq:Rf}) we can see that the dynamics of the field will not come from a standard kinetic term but from $R[f]$ and the potentials $\beta_i(\phi)U_i[\s]$.\\
\indent
It is also possible to identify the scalar longitudinal graviton mode $\phi$, with a standard canonical degree of freedom by adding an kinetic term to the action~(\ref{eq:action0}) as done in \cite{DeFelice:2017oym}. However we briefly discuss this scenario in Sec.~\ref{Chame} where we have shown its cosmological behavior

Now the action for the metric $f$, follows from the above and after some integration by parts, becomes
\begin{align}
\int \d^4x & \sqrt{-f}R[f]=\int \d^4x \sqrt{-\g}  \Bigl[\bar K(\phi,X)-\bar G_3(\phi,X)\DAlambert\phi \nonumber\\
&+\bar G_4(\phi,X)R+\bar G_{4X}(\phi,X)\left[(\DAlambert\phi)^2 -\phi_{\mu\nu}^2\right]\Bigr] 
\label{eq:action1}
\end{align}
with
\begin{align}
\bar K(\phi,X)&=-\frac{2X\left[B'^2(1-BX)-BB''(1-2BX) \right]}{B^2\sqrt{1-2BX}},\\
\bar G_3(\phi,X)&=\frac{(1-4BX)B'}{B\sqrt{1-2BX}},\\
\bar G_4(\phi,X)&=\sqrt{1-2BX},
\end{align}
which reduces to a class of Horndeski theory \cite{Horndeski:1974wa} and a form of higher dimensional gravity theories \cite{deRham:2010eu}, the quartic Galileon \cite{Nicolis:2008in,Deffayet:2009wt}. Therefore the full action becomes 
\begin{align}
\S=&\frac{M_g^2}{2}\int \d^4x \sqrt{-g}\Bigl[K(\phi,X)-G_3(\phi,X)\DAlambert\phi+G_4(\phi,X)R\nonumber\\
&+G_{4X}(\phi,X)[(\DAlambert\phi)^2 -\phi_{\mu\nu}^2]+m^2\sum_{i=0}^{4}\beta_i U_i[s]\Bigr] \, ,
\label{eq:action2}
\end{align}
where we define\\
\begin{eqnarray}
K(\phi,X)&=&\epsilon \al\bar K(\phi,X)  \, ,  \\     
G_3(\phi,X)&=& \epsilon \al\bar G_3(\phi,X) \, ,  \\  
G_4(\phi,X)&=&1+\epsilon  \al\bar G_4(\phi,X) \, ,  
\end{eqnarray}
with
\begin{eqnarray}
  \alpha &=& \frac{M_\f^2}{M_\g^2} \, , \\
  G_{iX} &=& \frac{\partial G_i}{\partial X} \, .
\end{eqnarray}

The action~(\ref{eq:action2}) can be easily generalized  by performing  a  transformation $f_{\mu\nu}=A(\phi)g_{\mu\nu}+B(\phi)\phi_{\mu}\phi_\nu$ on it. 

In the limit $X \ll 1$, the action becomes 
\begin{align}
\S=\frac{M_g^2}{2}\int \d^4x \sqrt{-g} \left[(1+\epsilon \alpha)R + m^2\sum_{i=0}^{4}\beta_i U_i[s] \right]
\end{align}
from  which we decide to define Planck mass as 
\begin{align}
M_g^2(1+\epsilon \alpha)\equiv \Mpl^2, \,\ \mbox{which gives} \,\  M_g^2  +\epsilon  M_f ^2 =\Mpl^2
\end{align}

\section{Cosmology}
\label{Cosmologysec}
\subsection{Field equations}

In order to incorporate the above theory in cosmological framework we have to introduce the matter and radiation sector in the action 
~(\ref{eq:action0}) and we assume here $\beta$ is independent of $\phi$. 
Next, we consider a flat Friedmann-Lema\^{i}tre-Robertson-Walker (FLRW) metric of the form
\begin{equation}
\d s^2=\g_{\mu\nu}\d x^\mu \d x^\nu=-N(t)^2 \d t^2 + a(t)^2\delta_{i j}\d x^i \d x^j \, ,
\end{equation}
where $t$  is the cosmic time and $x^i$ are the comoving spatial coordinates, $N(t)$ is the lapse function, and $a(t)$ is the scale factor. In FLRW geometry, $\phi$ becomes a function of $t$ only and thus $X=\frac{1}{2}\dot\phi^2(t)/N^2$.
To calculate the field equations we start with action (\ref{eq:action0}) including the matter action and vary it with respect to (w.r.t) $N(t)$ and $a(t)$ respectively, and set $N=1$, we obtain the Friedmann and Raychaudhuri equations respectively as 
\begin{align}
3H^2\Mpl^2&=\rho_\m+\rho_\r+\rho_\phi \, ,
\label{eq:hubble}\\
(2\dot H+3H^2)\Mpl^2 &=-\frac{\rho_\r}{3}-p_\phi \, ,
\label{eq:raycdhreqn}
\end{align}
where $\rho_\m$ and $\rho_\r$ are the matter and radiation energy densities. $\rho_\phi$ and $p_\phi$ are the energy density and pressure of the scalar field which are given by  
\begin{eqnarray}
\frac{\rho_\phi}{M_g^2}&=&\beta+\frac{\gam}{\sqrt{1-2 B X}} \nn \\&& +3 \al\ep\left(1-\frac{1}{(1-2 B X)^{3/2}}\right) H^2 \, ,~~~
\label{eq:rho}\\
\frac{p_\phi}{M_g^2}&=&-\bet-\gam\sqrt{1-2 B X}+\ep\al\Bigg\{\frac{\(B\dot X+X B_{\phi }\dot\phi\) 2H}{(1-2 B X)^{3/2}}\nn\\ &&+ \left(2 \dot{H}+3 H^2\right)\left(\frac{1}{\sqrt{1-2 B X}}-1\right)\Bigg\}\, .
\label{eq:pphi}
\end{eqnarray}
where
\begin{eqnarray}
\beta&=& -\frac{m^2}{2}\left(\beta_0+3\beta_1+3\beta_2+\beta_3\right)\, ,\\
\gamma&=&-\frac{m^2}{2}\left(\beta_1+3\beta_2+3\beta_3+\beta_4\right) \, .
\end{eqnarray}
Notice that with this definition, we have in FLRW background
\begin{align}
m^2\sum_{i=0}^{4}\beta_i U_i[s]=-2\beta-2\gamma \sqrt{1-2BX}
\end{align}
and therefore $\beta+\gamma$ plays the same role as a cosmological constant in the limit $X\ll 1$. Also From Eq.~(\ref{eq:rho}) and (\ref{eq:pphi}) one can see that in the absence of $R[f]$, we have $\Lambda$CDM model iff $\beta_1+3\beta_2+3\beta_3+\beta_4=0$ ($\gamma=0$).


The equation of motion of the scalar field is obtained by varying the action w.r.t the scalar field $\phi$ and is given by
\begin{eqnarray}
&&\bigg\{3\epsilon \al H^2(1+4BX)-\gam(1-2BX)\bigg\}\left(B\ddot\phi+XB_\phi\right)\nonumber\\
&&+(1-2BX)\bigg\{\epsilon \al(2\dot H+3H^2)-\gam(1-2BX)\bigg\}3BH\dot\phi \nn \\ && =0 \, .
\label{eq:kg}
\end{eqnarray}

\subsection{Dynamical System and fixed point analysis}

To obtain a dynamical system, we define the following dimensionless parameters
\begin{eqnarray}
x &=&\frac{\beta}{3H^2} \, ,
\label{eq:x}\\
y &=& \frac{\alpha^{1/3}}{(1-2BX)^\frac{1}{2}} \, ,
\label{eq:y}\\
\Omega_\r &=& \frac{\rho_\r}{3H^2\Mpl^2} \, , 
\label{eq:omr}\\
\Omega_\m &=& \frac{\rho_\m}{3H^2\Mpl^2} \, ,
\label{eq:omm}\\
\Omega_\phi &=& \frac{\rho_\phi}{3H^2\Mpl^2} \,
\label{eq:omphi}
\end{eqnarray}

Using the dimensionless parameters $x$ and $y$ we can define fractional energy density for the scalar field
\begin{equation}
\Om_\phi=\frac{x-\ep y^3 +\al \ep +\frac{\sigma x y}{\alpha^\frac{1}{3}}}{1+ \alpha \ep} \, ,
\end{equation}
where $\sig=\gam/\bet$. Also from Eq.~(\ref{eq:hubble}) we have $\Om_\m+\Om_\r+\Om_\phi=1$. So we can form a dynamical sysytem with three variables among the five variables defined in the Eqs.~(\ref{eq:x})-(\ref{eq:omphi}). We choose the first three variables (Eqs.~(\ref{eq:x})-(\ref{eq:omr})). The dynamical system is given below,
\begin{widetext}
\begin{eqnarray}
\label{eq:dynx}
\frac{\d x}{\d N} &=& x\Bigg\{3\al^{1/3} \sig^2 x^2+\left(3\ep\al^{1/3} y^3-y(2\ep\al+\sig x)\right)\left((1+\ep\al)\Om_\r+3(1-x)\right)-6\ep\al^{2/3}\sig x y^2+3\al y^4\Bigg\} \nn \\ && \Bigg / \Bigg\{y\Bigg(-2\ep\al-\sig x-\ep\al^{2/3}\sig xy+\al y^3+3\ep\al^{1/3}y^2\Bigg)\Bigg\} \, ,\\
\frac{\d y}{\d N} &=& -\frac{\(y^2-\al^{2/3}\)\(3x\left(\ep\al^{1/3} y^2-\sig\right)-y^2\ep\al^{1/3}(1+\ep\al)\Om_\r\)}{y\(-2\ep\al-\sig x-\ep\al^{2/3}\sig x y+\al y^3+3 \ep\al^{1/3}y^2\)}\, ,
\label{eq:dyny}\\
\frac{\d \Om_\r}{\d N} &=& \Om_\r\Bigg\{3\al^{1/3}\sig^2 x^2-2 \ep\al^{2/3}\sig xy^2-\al y^4+\Big(3\ep\al^{1/3}y^3-y(2\ep\al+\sig x)\Big)\Big((1+\ep\al)\Om_\r-(1+3x)\Big)\Bigg\}\nn \\ && \Bigg /\Bigg\{y\left(-2 \ep\al-\sig x-\ep\al^{2/3}\sig xy+\al y^3+3\ep\al^{1/3} y^2\right)\Bigg\} \, , 
\label{eq:dynomr}\\ 
\end{eqnarray}
\end{widetext}
where $N=\ln a$.

In terms of the dimensionless variables the effective and field's equation of states are given by
\begin{widetext}
\begin{eqnarray}
w_{\rm eff} &=& \frac{3\sig x^2\(\al^{1/3}\sig+y\)+xy\(6\epsilon \al-\sig(1+\ep\al)\Om_\r-9\epsilon \al^{1/3}y^2-3\epsilon \al^{2/3}\sig y\)+\ep\al^{1/3}(1+\ep\al)y\(3y^2-2 \al^{2/3}\)\Om_\r}{3y\(-2 \epsilon \al-\sig x-\epsilon \al^{2/3}\sig xy+ \al y^3+3\epsilon \al^{1/3} y^2\)} \, , ~~~~
 \label{eq:weff}\\
  w_\phi &=& \Bigg\{\al^{1/3}\left(1+\ep\al\right)\Bigg(-3 \sig  x^2 \left(\al^{1/3} \sig +y\right)+\epsilon x \left(9 \al^{1/3} y^3+\al^{2/3} \sig  y^2 (3-\Om_\r)-\al y(6-\sig\Om_\r)\right) \nn \\ && +\al y(y^3-3\al^{1/3} y^2+2\al) \Om_\r\Bigg)\Bigg\}\Bigg/\Bigg\{3 y \left(-2 \epsilon \al-\sig  x-\epsilon \al^{2/3} \sig  x y+\al y^3+3 \epsilon \al^{1/3} y^2\right) \Big(\epsilon \al^{1/3} (y^3-\al) \nn \\ && -x \left(\al^{1/3}+\sig  y\right)\Big)\Bigg\} \, .~~
 \label{eq:wphi}
\end{eqnarray}
\end{widetext}
One interesting thing to notice here is that the dynamical system and the equation of states are independent of the form of $B(\phi)$. In other words, the background cosmological dynamics is independent of the form of $B(\phi)$.
\begin{widetext}
\begin{table}[!htb]
\centering
\begin{tabular}{|p{1cm}||p{1.3cm}||p{1.4cm}||p{1cm}||p{1cm}||p{1cm}||p{1cm}||p{1.4cm}||p{2.5cm}||p{2.7cm}||}
\hline
\mbox{Pts.} & $x$ & $y$ & $\Om_\r$ & \mbox{$\Om_\phi$}& \mbox{$\Om_\m$}& $w_{\phi}$ & $w_{\rm eff}$ & Eigen Values & \mbox{Nature of Stability} \\
\hline
$P_R$ & 0 & $\al^{1/3}$ & $1$ & 0 & 0 & -1 & $1/3$ & 1, 2, 4 & unstable \\
\hline
$P_M$ & 0 & $\al^{1/3}$ & 0  & 0 & 1 & -1 & 0 & -1, 0, 3 & saddle\\
\hline
$P_{dS}$  & $\frac{1+\epsilon \al}{1+\sig}$ & $\al^{1/3}$ & 0  & 1 & 0 & -1 &-1 & -6, -4, -3 & stable\\
\hline
L & 0 & $y\neq 0$ & 0  & $\frac{\ep(\al-y^3)}{1+\ep\al}$ & $\frac{1+\epsilon y^3}{1+\ep\al}$ & 0 & 0 & -1, 0, 3 & saddle\\
\hline
\end{tabular}
\begin{tabular}{p{4.3cm} p{10.6cm}}
Special point only for $\epsilon=1$&\\
\end{tabular}
\begin{tabular}{|p{1cm}||p{1.3cm}||p{1.4cm}||p{1cm}||p{1cm}||p{1cm}||p{1cm}||p{1.4cm}||p{2.5cm}||p{2.7cm}||}
\hline$ P_S $ & 1 & $\frac{\sqrt{\sig}}{\al^{1/6}}$ & 0  & 1 & 0 & -1 &-1 & -4, -3, -3 & stable\\
\hline
\end{tabular}
\caption{Critical points, their nature and stability.}
\label{tab:dyn}
\end{table}
\end{widetext}

In Table~\ref{tab:dyn}, we have listed all fixed points of our model. Among these points we can see that the first three points correspond to standard cosmology with $P_R$ being radiation era, $P_M$ the matter era and $P_{dS}$ the $dS$ universe. They all belong to the same subspace $y=\alpha^{1/3}$ which corresponds to $\dot \phi=$constant. We also notice that we could impose $\epsilon=-1$.

To these three critical points, we have an additional critical line $P_L$ corresponding to $\phi$MDE, which is an era dominated by the scalar field but behaving like a matter dominated era. It is standard in various theories with non-minimally coupled scalar field or $f(R)$ theory \cite{Amendola:2006we}.

And finally, the special case $\epsilon=+1$ gives rise to an additional non-trivial dS point. In fact, $P_{dS}$ could be reached with a trivial scalar field $\dot \phi$ constant from radiation era, therefore a $\Lambda$CDM model. In order to have a non-trivial dynamics, we need to reach the point $P_S$ which means that $y$ would change from the radiation era. Therefore, if one look to some new physics, it is wise to choose $\epsilon=+1$.
\begin{figure}[H]
\includegraphics[scale=.75]{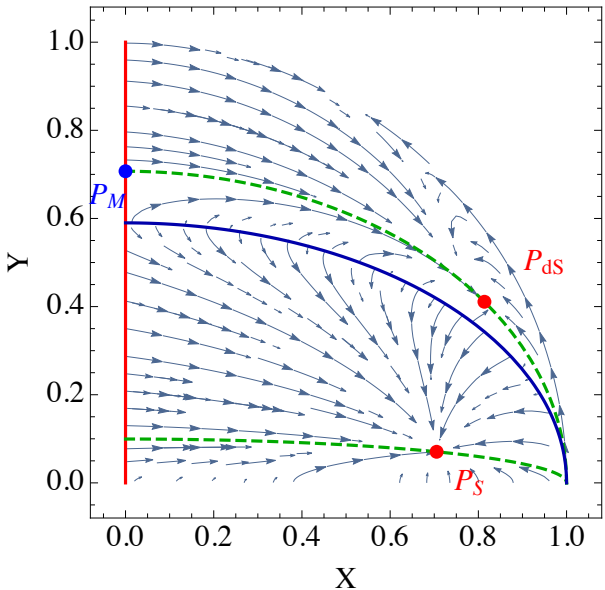}
\caption{Phase space in the Poincar\'e sphere describing the dynamics of the system in the late universe $(\Omega_r=0)$ for $\epsilon=1$, $\alpha=1$ and $\sigma=0.01$} 
\label{fig:phasespace}
\end{figure}
We see from Fig.~\ref{fig:phasespace}, the very interesting phase space of this system.We have reduced the system to $\Omega_r=0$ which corresponds to late universe. Also we have represented the phase space in the Poincar\'e coordinates $(X,Y)$ 
\begin{align}
X=\frac{x}{\sqrt{1+x^2+y^2}}\\
Y=\frac{y}{\sqrt{1+x^2+y^2}}
\end{align}
which permits to have access to the full phase space by compactifying the variables over a sphere. This system is reduced to $Y>0$ because of the definition of this variable and $X>0$. The case $X<0$ being similar.

We have two de Sitter points belonging to invariant submanifolds represented in green dashed line. It is interesting to notice that each one defines an invariant subspace separated by the blue line representing an other invariant submanifold. Therefore we have 2 separated subspaces with each one an attractor point. The matter point belongs to the upper subspace and therefore we have a trajectory $P_M\rightarrow P_{dS}$ corresponding 
to a trivial cosmology ($\dot \phi$ constant). The interesting attractor would be $P_S$ for which $\phi(t)$ would have a non-trivial evolution during the expansion of the universe. Unfortunately, this point is decoupled from $P_M$ and hence we could not have a standard evolution of matter era followed by de Sitter universe. But as we have noticed previously, the model has a critical line for $X=0$ which represents a $\phi$MDE, therefore we could have a $\phi$MDE followed by $P_S$. This would be compatible if we start from a critical point close to $P_M$ and therefore reduce the presence of the scalar field during the matter era. That would be viable if $\alpha$ is sufficiently small.

\begin{figure}[H]
\includegraphics[scale=.75]{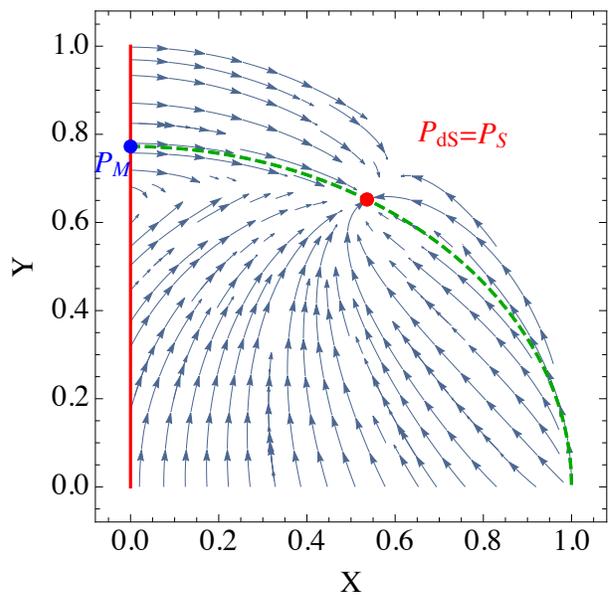}
\caption{Phase space in the Poincar\'e sphere describing the dynamics of the system in the late universe $(\Omega_r=0)$ for $\epsilon=1$, $\alpha=1$ and $\sigma=1$} 
\label{fig:phasespace2}
\end{figure}

Notice finally that if $\sigma=\alpha$, the two de Sitter points merge into one critical point as seen in Fig.~\ref{fig:phasespace2}

\begin{figure}[H]
\includegraphics[scale=.75]{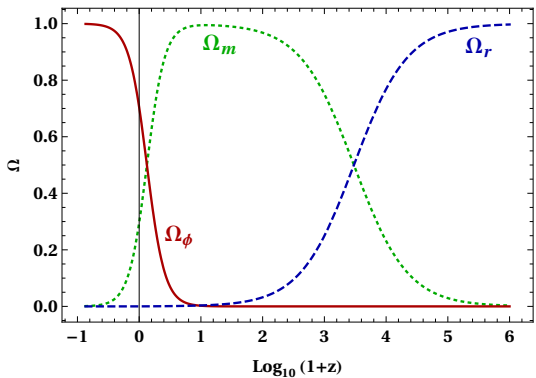}
\caption{Green (dotted), blue (dashed) and red (solid) lines represent evolution of the fractional energy density of matter, radiation and scalar field respectively for parameters $\sig=1$ and $\al=1$.} 
\label{fig:density}
\end{figure}

\begin{figure}[H]
\includegraphics[scale=.75]{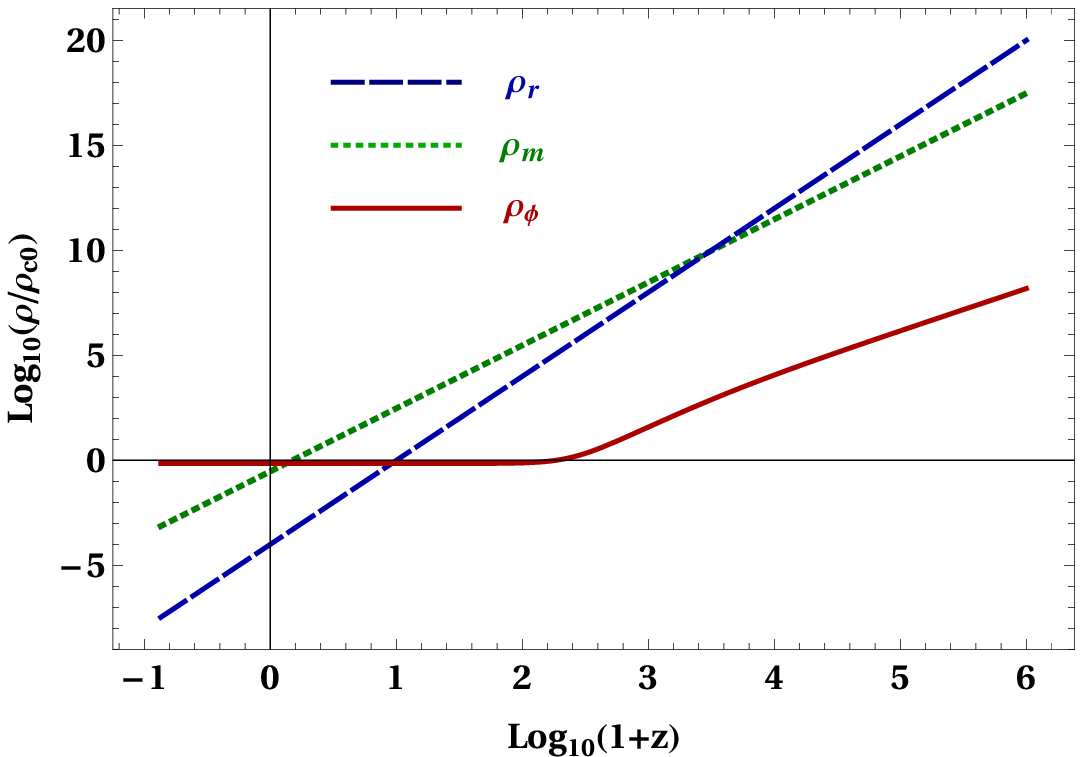}
\includegraphics[scale=.75]{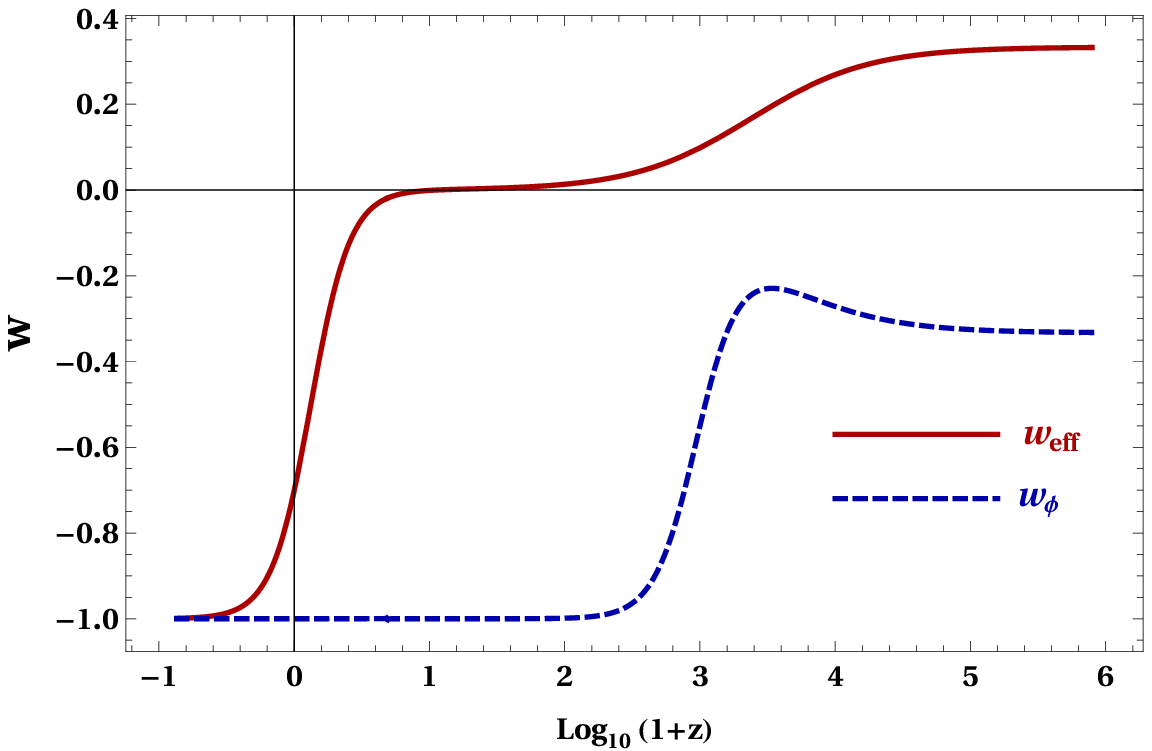}
\caption{{\bf Up:} Green (dotted), blue (dashed) and red (solid) lines represent evolution of the energy density of matter, radiation and scalar field respectively. $\rho_{\rm c0}$ is the present critical density of the Universe. {\bf Bottom:} Evolution of equation of states are shown. Red (solid) line represents effective equation of state and blue (dotted) line represents scalar field equation of sate. For both the plots $\sig=1$ and $\al=1$.} 
\label{fig:rho}
\end{figure}

In Fig.~\ref{fig:density} we have shown the evolution history of the universe in terms of the fractional energy density $\Omega$. Same is shown in the Fig.~\ref{fig:rho} in terms of energy density $\rho$ and equation of state $w$. The figures show that the viable cosmological evolution of the cosmological parameters can be achieved in the scenario under consideration.

\section{Tensor Perturbations}
\label{TenPert}
In this section, we will study the tensor perturbations of this model. Considering the previous background, we decompose the metric in the ADM form
\begin{align}
{\rm d}s^2=-N^2 {\rm d}t^2+\gamma_{ij}\Bigl({\rm d}x^i+N^i {\rm d}t\Bigr) \Bigl({\rm d}x^j+N^j {\rm d}t\Bigr)
\end{align}
Following the standard process \cite{Maldacena:2002vr}, the metric is expanded around the FLRW metric by considering
\begin{align}
N &=1\\
N_i &=0\\
\gamma_{ij} &=a^2 \Bigl(\delta_{ij}+h_{ij}+\frac{1}{2}h_{ik}h^k_{~j} \Bigr)
\end{align}
With these definitions, the action is expanded to the second order and the quadratic action is found to be
\begin{align}
\mathcal{S}^{(2)}_T=\int {\rm d}^4 x a^3 \mathcal{G}_T \Bigl[ \dot{h}_{ij}^2-\frac{c_T^2}{a^2}(\vec \nabla h_{ij})^2\Bigr]
\end{align}
where 
\begin{align}
\mathcal{G}_T=1+\epsilon \alpha^{2/3} y
\end{align}
and squared sound speed is given by
\begin{align}
c_T^2=\frac{1+\epsilon \alpha^{4/3}/y}{1+\epsilon \alpha^{2/3} y}
\end{align}
These formulas are consistent with what was previously found in \cite{Kobayashi:2011nu}. Notice that $c_T^2=1$ for $y=\alpha^{1/3}$ which corresponds to an invariant submanifold. Therefore during the evolution either $c_T>1$ or $c_T<1$ but can't cross the line $c_T=1$. 

The detection of the binary neutron star merger GW170817 and its associated electromagnetic counterparts have put very strong constraints on $c_T$ \cite{TheLIGOScientific:2017qsa,Monitor:2017mdv}.
\begin{align}
1-3.10^{-15}\leq \frac{c_T}{c}\leq 1+7.10^{-16}
\end{align}

Considering that our model converges to $P_{dS}$, we would have $y\simeq \alpha^{1/3}$ today, which gives $c_T\simeq 1$. But if we consider the model to converge to $P_{S}$, we would have $y\simeq \sqrt{\sigma}/\alpha^{1/6}$ today and therefore
\begin{align}
c_T^2=\frac{1+\alpha^{3/2}/\sqrt{\sigma}}{1+\sqrt{\alpha \sigma }}
\end{align}
We will have $c_T\simeq 1$ if we consider $\alpha\simeq \sigma$ and therefore if $P_S=P_{dS}$. We see therefore from this analysis that the second attractor ($P_S$) is excluded except if extreme fine-tuning of the parameters. We can conclude that only the standard cosmology $P_R\rightarrow P_M\rightarrow P_{dS}$ is consistent with latest constraints from gravitational wave experiment.
\section{DYNAMICS IN CASE OF $\phi$ DEPENDENT $\beta$}
\label{Chame}
In this section we promote the coefficient of the potential  $U_i$ as a function of the scalar field $\phi$  via, $\beta_i = \beta_i(\phi)$ and also add a canonical kinetic term for the scalar field. By doing so, we promote environment dependency of the mass term as discussed in ref.(\cite{DeFelice:2017oym}) where one can implement chameleon mechanism. Here we only briefly discuss its cosmological behavior By this inclusion our action takes the following form 
\begin{align}
\S_c=&\frac{M_g^2}{2}\int \d^4x \sqrt{-g} R[g] +\epsilon\frac{M_f^2}{2}\int \d^4x \sqrt{-f} R[f]  \nonumber\\
&+ \frac{M_g^2}{2}m^2 \int \d^4x \sqrt{-g}\sum_{i=0}^{4}\beta_i(\phi) U_i[\s] \nonumber\\
 &- \frac{1}{2}\int \d^4x \sqrt{-g} g^{\mu\nu}\partial_\mu\phi \partial_\nu\phi \, ,
\label{eq:actionchem}
\end{align}
Varying the action  w.r.t $N(t), a(t)$ and $\phi(t)$, we obtain Friedmann equation, evolution equation and Klein-Gordon equation respectively. Writing  in a compact way we respectively found 
\begin{widetext}
\begin{eqnarray}
\rho_{\rm\phi} & = & 3 H^2 \Mpl^2 + \frac{\Mpl^2 \beta}{1+\alpha\epsilon}+\frac{1}{2}\dot{\phi}^2 + \frac{ \Mpl^2 \gamma }{(1+\alpha  \epsilon ) \sqrt{1-B \dot{\phi}^2}}-\frac{3 H^2\Mpl^2 \( 1+ \frac{\alpha \epsilon}{\(1-B\dot{\phi}^2 \)^{\frac{3}{2}}}\)}{1+\alpha\epsilon},
 \label{eq:rhophimod}\\
p_\phi & = & -\Mpl^2 \(2 \dot{H}+3 H^2\)-\frac{\Mpl^2 \beta}{1+\alpha\epsilon}+\frac{1}{2}\dot{\phi}^2 +\frac{\Mpl^2 \gamma}{\(1+\alpha \epsilon \)\(1-B\dot{\phi}^2\)^{\frac{3}{2}}}-\frac{\Mpl^2\gamma B^2\dot{\phi}^4}{\(1+\alpha \epsilon \)\(1-B\dot{\phi}^2\)^{\frac{3}{2}}}-\frac{2 \Mpl^2 \gamma}{\(1+\alpha\epsilon\)\sqrt{1-B\dot{\phi}^2}}\nonumber\\
&& +\(2 \dot{H}+3 H^2\)\(\frac{\Mpl^2}{1+\alpha\epsilon}+\frac{\Mpl^2 \alpha\epsilon}{\(1+\alpha \epsilon \)\sqrt{1-B\dot{\phi}^2}}\)+\frac{2\Mpl^2 B H \alpha\epsilon\dot{\phi}\ddot{\phi}}{\(1+\alpha \epsilon \)\(1-B\dot{\phi}^2\)^{\frac{3}{2}}}\label{eq:pphimod},\\
&& \bar{G_4}^4 \Mpl^2 \( \bar{G_4} \beta'+\gamma'\) +3 \G4^2 H \dot{\phi} \left\lbrace \G4^2\( \G4 + B \Mpl^2 \gamma\)+\(\G4^3 -\Mpl^2 B \( 2 \dot{H}+3 H^2\) \alpha\epsilon \) \right\rbrace \nonumber\\
 && \left\lbrace \G4^5 + B \G4^2 \Mpl^2 \gamma + \( \G4^5 +3 B \(-3 +2 \G4^2\) H^2 \Mpl^2 \)\alpha \epsilon \right\rbrace \ddot{\phi}=0.\label{eq: KGmod}
\end{eqnarray}
\end{widetext}
where $\beta=\beta(\phi),\gamma=\gamma(\phi)$ defined the same way as above and $'$ denotes  the derivative w.r.t $\phi$ and $\G4=\sqrt{1-B\dot{\phi}^2}$. 

\subsection{dynamical system}
To understand the behavior of the above system we define some additional dimensionless  variables($x$ and $y$ are as defined above, $\gamma=\beta \sigma$ and $\gamma'=\beta'\sigma$) \\
\begin{eqnarray}
z^2&=& \frac{\dot{\phi}^2}{6 H^2 \Mpl^2}\\
\k&=& \Mpl\frac{\beta'}{\beta}\\
\Gamma_\phi &=&\frac{\beta'' \beta}{\beta'^2}
\end{eqnarray}
Using dimensionless variables we obtain \\
\begin{eqnarray}
w_\phi &=& -\frac{\alpha^{1/3} \left(\alpha  \epsilon +1\right) \big(A_1+A_2+A_3+x (A_4+A_5+A_6)\big)}{A_7~D}\nonumber\\
\end{eqnarray}
\begin{eqnarray}
w_{eff}&=& -1 + \frac{2 y~ B_1~\epsilon  \left(y^2-\alpha ^{2/3}\right)-B_2~B_3}{3 \sqrt{6}~ D~y}
\end{eqnarray}

Using these variables we get the constraint equation \\
\begin{eqnarray}
\Om_\phi +\Om_r + \Om_m=1
\end{eqnarray}
with 
\begin{eqnarray}
\Om_\phi= \frac{\alpha  \epsilon +\frac{\sigma  x y}{\alpha^{1/3}}+x-\epsilon y^3 +\alpha  z^2 \epsilon +z^2}{\alpha  \epsilon +1}.
\end{eqnarray}
We have the evolution equations for the the dynamical variables as
\begin{eqnarray}
\frac{dx}{dn}&=& \frac{x}{\sqrt{6}}\Bigg \lbrace  6z\kappa_\phi +\frac{\Big(\left( 2y \left(y^2-\al^{2/3}\right)\epsilon A - B C\right)\Big)}{y~D} \Bigg\rbrace\nonumber\\ 
\label{eq:dxdnmod}
\\
\frac{dy}{dn}&=&\left(\frac{y^2}{\alpha ^{2/3}}-1\right)\frac{E +F +G}{\sqrt{6}~D}\label{eq:dydnmod}\\
\frac{dz}{dn}&=&\frac{z (P-Q+R+S+T)}{2 \sqrt{6}~D~y}\label{eq:dzdnmod}\\
\frac{d\Om_r}{dn}&=& \Omega_r \left(\frac{M-N~O}{\sqrt{6}D ~y}-4\right)\label{eq:domdnmod}\\
\frac{d \kappa_\phi}{dn}&=& \sqrt{6} (\Gamma_\phi -1) \kappa_\phi ^2 z\label{eq:dka}
\end{eqnarray}
with $A_1, A, B $ etc. are given in appendix~A(\ref{appendA})\\
\begin{figure}[H]
\includegraphics[scale=.75]{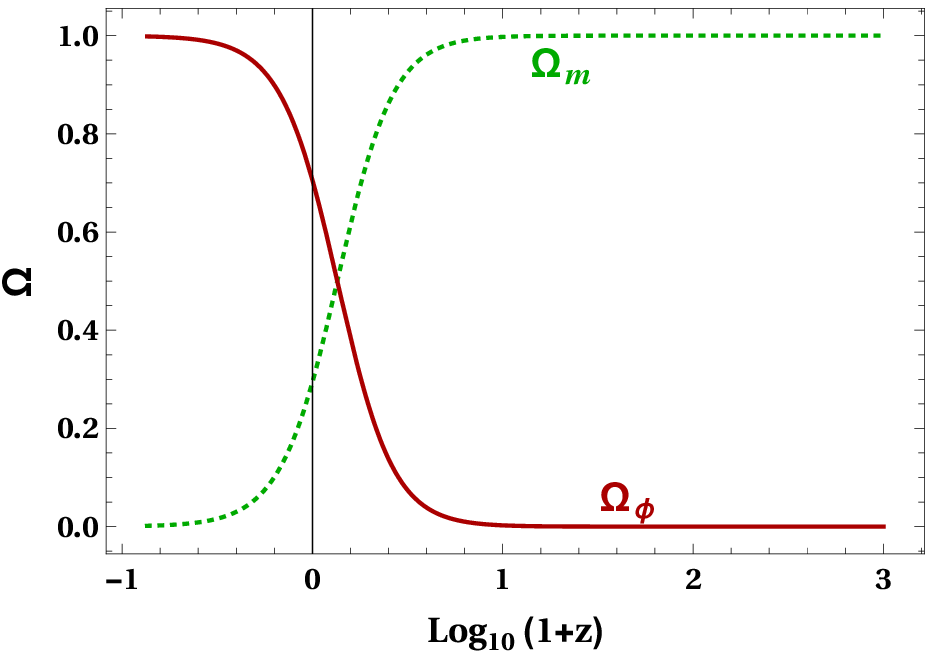}
\includegraphics[scale=.75]{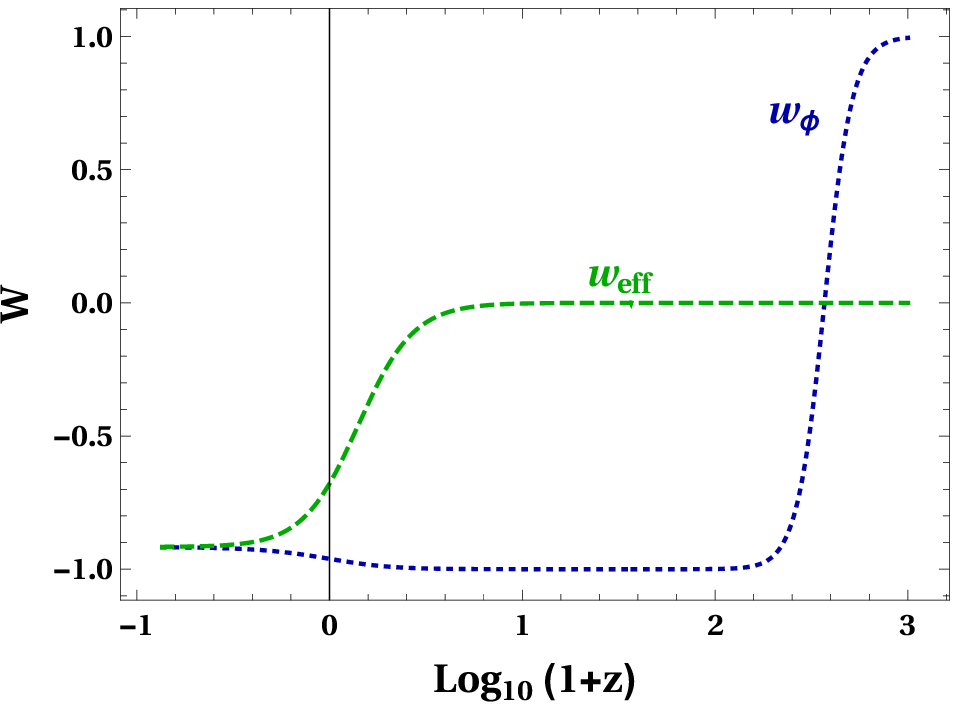}
\caption{{\bf Up:} Green (dotted) and red (solid) lines represent evolution of the fractional energy density of matter and scalar field respectively. {\bf Bottom:} Evolution of equation of states are shown. Green (dashed) line represents effective equation of state and blue (dotted) line represents scalar field equation of sate. For both the plots $\sig=1$, $\al=0.12$ and $\lam=0.5$.} 
\label{fig:Om_and_w_mod}
\end{figure}
\subsection{Numerical analysis}
\label{Numerical_mod}
To study the cosmological late time behavior of our present system, we further consider $\Om_r=0$, as we are concerned with late time only and  choose a specific form for the coefficient $\beta$ as\\
\begin{eqnarray}
\beta(\phi)= \beta_M e^{\lambda\phi/\Mpl}, 
\end{eqnarray}
with $\beta_M$ is some constant with dimension of $\Mpl^2$ and $\lambda$ is a dimensionless constant. These further simply the system by making $\kappa_\phi=\lambda$ and keep only three relevant dynamical equations(\ref{eq:dxdnmod},~\ref{eq:dydnmod} and \ref{eq:dzdnmod}). Here we only focus on whether the scalar dof gives an accelerated expansion and leave a more systematic analysis of all critical points for future project. The numerical results shows that scalar field, $\phi$ indeed gives a suitable dark energy behavior. It is very clear from Fig.\ref{fig:Om_and_w_mod} that the scalar field starts out from a kinetic regime and then enters the accelerated phase, So we can conclude that it has needed shows a viable dark energy behavior. 

Thus we can say that $\phi$ dependent $\beta$ case ( set up for implementation of chameleon mechanism) do not spoil the dark energy behavior, although it significantly alter its nature.  

\section{Conclusion}
\label{concl}
In this paper, we studied a specific limit of bi-gravity where we have focused on one additional scalar degree of freedom. We have shown that the framework, in this case, reduces to DBI class of theories where the mass term of the graviton splits into two parts, one behaving like a cosmological constant whereas the other looks  like $\sqrt{1-2BX}$. Considering this new model, we have studied the background evolution using the dynamical system approach. We have demonstrated the existence of two attractors with two basins of attraction separated by an invariant submanifold. One of the regions is relevant  to standard cosmology where the acceleration is provided by the effective cosmological constant generated by the mass of the graviton. The second attractor $P_S$ is associated with a new dynamics where the scalar field evolves in time such that $P_S$ can not be reached starting from radiation era thereby no viable cosmology in this case.
Finally, we have studied the scalar tensor perturbations in the model under consideration and showed that if the first attractor is consistent with observations at late time, the second attractor is  generally disfavored because the speed of propagation of gravitational waves deviates too much from current constraints, the requirement of $C_T$ to be extremely close to one give rise to large fine tuning in $\alpha$. As for  the first attractor, $P_R\to P_M \to P_{dS}$, dynamics  is consistent with latest constraints from gravitational wave experiment. 
Thus we conclude that focusing  only on one degree of freedom provided by the second sector of gravity, it is difficult  to accommodate all current constraints  without invoking large fine-tuning.\\
It is also clear that that this model exhibits a proper dark energy behavior even after its modification to consider environment dependency.

\section*{Acknowledgments}
The work of R. Gannouji is partially supported by DII-PUCV No 039.450/2017 and Fondecyt project No 1171384. M. W. Hossain is supported by the Ministry of Science, ICT \& Future Planning, Gyeongsangbuk-do and Pohang City. N. Jaman is thankful to Safia Ahmad and Bikash Ranjan Dinda for discussion. His work is funded by UGC, Govt. of India. 

\begin{widetext}
\section{Appendix A}
\label{appendA}

\begin{eqnarray}
A_1 &=& -3 \sigma  x^2 y \left(y^2-\alpha ^{2/3}\right) \left(\alpha^{1/3}\sigma +y\right)+\alpha  y^7 \Om_r \epsilon ^2-3 \alpha^{1/3}y^6 \epsilon  \left(\alpha  \Om_r \epsilon +3 z^2 (\alpha  \epsilon +1)\right)-\alpha ^{5/3} y^5 \Om_r \epsilon ^2,\\
A_2&=& 2 \alpha  y z^2 (\alpha  \epsilon +1) \left(\alpha  \Om_r \epsilon +3 z^2 (\alpha  \epsilon +1)\right)-2 \alpha ^{5/3} y^2 \epsilon  \bigg(\alpha  \Om_r \epsilon +(\Om_r-3) z^2 (\alpha  \epsilon +1)\bigg),\\
A_3&=& \alpha  y^4 \epsilon  \left(5 \alpha  \Om_r \epsilon +3 z^2 (\alpha  \epsilon +1)\right),\\
A_4&=& 9 \alpha^{1/3}y^6 \epsilon -\alpha ^{2/3} \sigma  y^5 \epsilon  \left(\Om_r+2 \sqrt{6} \kappa_\phi  z-3\right)-6 \alpha  y z^2 (\alpha  \epsilon +1)-6 \alpha ^{4/3} \sigma  z^2 (\alpha  \epsilon +1),\\
A_5&=& \alpha ^{4/3} \sigma  y^3 \epsilon  \left(\Om_r+2 \sqrt{6} \kappa \phi  z-3\right)+\alpha ^{2/3} y^2 \bigg(\alpha  \epsilon  \left(-\sigma  \Om_r-3 \sigma  z^2+2 \sqrt{6} \kappa_\phi  z+6\right)-3 \sigma  z^2\bigg),\\
A_6&=& y^4 \bigg(\alpha  \epsilon  \left(\sigma  \Om_r+3 \sigma  z^2-2 \sqrt{6} \kappa \phi  z-15\right)+3 \sigma  z^2\bigg),\\
A_7&=& 3 y \bigg\{x \left(\alpha^{1/3}+\sigma  y\right)+\alpha^{1/3} \bigg(\epsilon  \left(\alpha -y^3\right)+z^2 (\alpha  \epsilon +1)\bigg)\bigg\},\\
B_1 &=& -3 \sqrt{6} \alpha ^{4/3} \sigma  x+3 \alpha ^{2/3} y^2 \left(\sqrt{6} \alpha  \epsilon +\sigma  x \left(2 \kappa \phi  z+\sqrt{6}\right)\right)+6 \alpha  y z \left(\kappa \phi  x+\sqrt{6} z (\alpha  \epsilon +1)\right)-3 \sqrt{6} \alpha  y^4 \epsilon,\\
B_2&=& \sqrt{6} \left(2 \alpha  z^2 (\alpha  \epsilon +1)-\left(y^3-\alpha ^{2/3} y\right) \left(-2 \alpha  \epsilon -\sigma  x+3 \alpha^{1/3} y^2 \epsilon \right)\right),\\
B_3&=& -3 \alpha^{1/3} \sigma  x+y \bigg(\alpha  \Om_r \epsilon -3 x+\Om_r+3 z^2 (\alpha  \epsilon +1)+3\bigg)+3 \alpha ^{2/3} y^2 \epsilon,\\
A &=& -3 \sqrt{6} \alpha ^{4/3} \sigma  x+3 \alpha ^{2/3} y^2 \left(\sqrt{6} \alpha  \epsilon +\sigma  x \left(2 \kappa \phi  z+\sqrt{6}\right)\right)+6 \alpha  y z \left(\kappa \phi  x+\sqrt{6} z (\alpha  \epsilon +1)\right)-3 \sqrt{6} \alpha  y^4 \epsilon,\\
B&=&\sqrt{6} \left(2 \alpha  z^2 (\alpha  \epsilon +1)-\left(y^3-\alpha ^{2/3} y\right) \left(-2 \alpha  \epsilon -\sigma  x+3 \alpha^{1/3}y^2 \epsilon \right)\right),\\
C&=&-3 \alpha^{1/3}\sigma  x+y \left(\alpha \Omega_r\epsilon -3 x+\Omega_r+3 z^2 (\alpha  \epsilon +1)+3\right)+3 \alpha ^{2/3} y^2 \epsilon,\\
D&=& y \left(y^2-\alpha ^{2/3}\right) \left(-2 \alpha  \epsilon -\sigma  x-\alpha ^{2/3} \sigma  x y \epsilon +\alpha  y^3 \epsilon ^2+3 \alpha^{1/3}y^2 \epsilon \right)-2 \alpha  z^2 (\alpha  \epsilon +1) \left(\alpha ^{2/3} y \epsilon +1\right),\\
E&=& -3 \sqrt{6} \alpha ^{4/3} \sigma  x+6 \alpha ^{4/3} \kappa_\phi  \sigma  x y^3 z \epsilon +6 \alpha  y z \left(\kappa_\phi  x+\sqrt{6} z (\alpha  \epsilon +1)\right),\\
F&=& \alpha ^{2/3} y^2 \left(3 x \left(2 \kappa_\phi  z+\sqrt{6}\right) (\alpha  \epsilon +\sigma )+\sqrt{6} \alpha  \epsilon  (\alpha  \epsilon +1) \left(3 z^2-\Omega_r\right)\right),\\
G&=& \sqrt{6} \alpha  y^4 \epsilon  \Big(\left(\alpha  \epsilon +1\right) \left(\Omega_r+3 z^2\right)-3 x\Big)\\
S&=& 3 \sqrt{6} \alpha^{1/3}\sigma  x y^3 (2 \alpha  \epsilon +\sigma  x)+6 \sqrt{6} \alpha ^{4/3} \sigma  x \left(z^2 (\alpha  \epsilon +1)-1\right)+3 \sqrt{6} \alpha  y^7 \epsilon ^2,\\
P&=& 3 \sqrt{6} \alpha^{1/3}y^6 \epsilon  \Big(\alpha  \Om_r\epsilon -3 x+\Om_r+3 z^2 (\alpha  \epsilon +1)+3\Big)-3 \alpha ^{2/3} y^5 \epsilon  \left(\sqrt{6} \alpha  \epsilon +2 \sigma  x \left(\sqrt{6}-2 \kappa_\phi  z\right)\right),\\
Q&=& \alpha  y \Big\lbrace3 \sqrt{6} \sigma ^2 x^2+2 \sqrt{6} z^2 (\alpha  \epsilon +1) (\alpha  \Omega_r \epsilon -3 x+\Omega_r-3)-12 \kappa_\phi  x z+6 \sqrt{6} z^4 (\alpha  \epsilon +1)^2\Big\rbrace,\\
R&=& \alpha ^{2/3} y^2 \Big\lbrace\sqrt{6} \alpha  \epsilon  \bigg(\sigma  x \Omega_r+3 z^2 (\sigma  x-2)+6\bigg)+\sigma  x \left(-3 \sqrt{6} x+\sqrt{6} (\Omega_r+9)+3 \sqrt{6} z^2+12 \kappa \phi  z\right)-6 \sqrt{6} \alpha ^2 z^2 \epsilon ^2\Big\rbrace,\\
T&=&y^4 \Bigg[-\alpha  \epsilon  \Big\{x \left(\sqrt{6} (\sigma  \Om_r-9)+3 \sqrt{6} \sigma  z^2-12 \kappa_\phi  z\right)+3 \sqrt{6} \left(\Om_r-z^2+5\right)\Big\}\nonumber\\
&& \,\, +\sqrt{6} \sigma  x \left(3 x-\Om_r-3 z^2-3\right)+3 \sqrt{6} \alpha ^2 \epsilon ^2 \left(z^2-\Om_r\right)\Big],\\
M&=& 2 y \epsilon  \left(y^2-\alpha ^{2/3}\right) \Big\{-3 \sqrt{6} \alpha ^{4/3} \sigma  x+3 \alpha ^{2/3} y^2 \bigg(\sqrt{6} \alpha  \epsilon +\sigma  x \left(2 \kappa \phi  z+\sqrt{6}\right)\bigg)+6 \alpha  y z \left(\kappa_\phi  x+\sqrt{6} z (\alpha  \epsilon +1)\right)-3 \sqrt{6} \alpha  y^4 \epsilon \Big\},\nonumber
\\
\\
N&=& \sqrt{6} \bigg\{2 \alpha  z^2 (\alpha  \epsilon +1)-\left(y^3-\alpha ^{2/3} y\right) \left(-2 \alpha  \epsilon -\sigma  x+3 \alpha^{1/3} y^2 \epsilon \right)\bigg\},\\
O&=& -3 \alpha^{1/3} \sigma  x+y \bigg(\alpha  \Om_r \epsilon -3 x+\Om_r+3 z^2 (\alpha  \epsilon +1)+3\bigg)+3 \alpha ^{2/3} y^2 \epsilon.
\end{eqnarray}
\end{widetext}


\begin{thebibliography}{150}
\bibitem{Brans:1961sx} 
  C.~Brans and R.~H.~Dicke,
  \href{\doi/10.1103/PhysRev.124.925}{Phys.\ Rev.\  {\bf 124}, 925 (1961).}
\bibitem{Lovelock:1972vz} 
  D.~Lovelock,
  \href{\doi/10.1063/1.1666069}{J.\ Math.\ Phys.\  {\bf 13}, 874 (1972)}.

\bibitem{massive} 
  K.~Hinterbichler,
  \href{\doi/10.1103/RevModPhys.84.671}{Rev.\ Mod.\ Phys.\  {\bf 84}, 671 (2012)}
  [\href{\arxiv/arXiv:1105.3735}{arXiv:1105.3735} [hep-th]];\,\ 
   C.~de Rham,
  \href{\doi/10.12942/lrr-2014-7}{Living Rev.\ Rel.\  {\bf 17}, 7 (2014)}
  [\href{\arxiv/arXiv:1401.4173}{arXiv:1401.4173} [hep-th]];
 M.~Sami and R.~Myrzakulov,
  \href{\doi/10.1142/S0218271816300317}{Int.\ J.\ Mod.\ Phys.\ D {\bf 25}, no. 12, 1630031 (2016)}
  [\href{\arxiv/arXiv:1309.4188}{arXiv:1309.4188} [hep-th]]; S.~Nojiri, S.~D.~Odintsov and M.~Sami,
  \href{\doi/10.1103/PhysRevD.74.046004}{Phys.\ Rev.\ D {\bf 74}, 046004 (2006)}
  [\href{\arxiv/hep-th/0605039}{hep-th/0605039}]. 




\bibitem{Fierz:1939ix} 
  M.~Fierz and W.~Pauli,
  \href{\doi/10.1098/rspa.1939.0140}{Proc.\ Roy.\ Soc.\ Lond.\ A {\bf 173}, 211 (1939)}.
\bibitem{Vainshtein:1972sx} 
  A.~I.~Vainshtein,
  \href{\doi/10.1016/0370-2693(72)90147-5}{Phys.\ Lett.\  {\bf 39B}, 393 (1972)}.

\bibitem{Dvali:2000hr} 
  G.~R.~Dvali, G.~Gabadadze and M.~Porrati,
  \href{\doi/10.1016/S0370-2693(00)00669-9}{Phys.\ Lett.\ B {\bf 485}, 208 (2000)}
  [\href{\arxiv/hep-th/0005016}{hep-th/0005016}].

\bibitem{deRham:2010kj} 
  C.~de Rham, G.~Gabadadze and A.~J.~Tolley,
  \href{\doi/10.1103/PhysRevLett.106.231101}{Phys.\ Rev.\ Lett.\  {\bf 106}, 231101 (2011)}
  [\href{\arxiv/arXiv:1011.1232}{arXiv:1011.1232} [hep-th]].
\bibitem{deRham:2010ik} 
  C.~de Rham and G.~Gabadadze,
  \href{\doi/10.1103/PhysRevD.82.044020}{Phys.\ Rev.\ D {\bf 82}, 044020 (2010)}
  [\href{\arxiv/arXiv:1007.0443}{arXiv:1007.0443} [hep-th]].

\bibitem{DeFelice:2012mx}
  A.~De Felice, A.~E.~Gumrukcuoglu and S.~Mukohyama,
  \href{\doi/10.1103/PhysRevLett.109.171101}{Phys.\ Rev.\ Lett.\  {\bf 109} (2012) 171101}
  [\href{\arxiv/arXiv:1206.2080}{arXiv:1206.2080} [hep-th]].

\bibitem{deRham:2012kf}
  C.~de Rham and S.~Renaux-Petel,
  \href{\doi/10.1088/1475-7516/2013/01/035}{JCAP {\bf 1301} (2013) 035}
  [\href{\arxiv/arXiv:1206.3482}{arXiv:1206.3482} [hep-th]].

\bibitem{Hassan:2011zd}
  S.~F.~Hassan and R.~A.~Rosen,
  \href{\doi/10.1007/JHEP02(2012)126}{JHEP {\bf 1202} (2012) 126}
  [\href{\arxiv/arXiv:1109.3515}{arXiv:1109.3515} [hep-th]].


\bibitem{Darkenergy} 
  E.~J.~Copeland, M.~Sami and S.~Tsujikawa,
  \href{\doi/10.1142/S021827180600942X}{Int.\ J.\ Mod.\ Phys.\ D {\bf 15}, 1753 (2006)}
  [\href{\arxiv/hep-th/0603057}{hep-th/0603057}];
  M.~Sami,
  \href{\doi/10.1007/978-3-540-71013-4\_8}{Lect.\ Notes Phys.\  {\bf 720}, 219 (2007)};	
M.~Sami,
Why is Universe so dark? 
New Adv.Phys. 10 (2016) 77-105 [physics.pop-ph];

  S.~M.~Carroll,
  \href{\doi/10.12942/lrr-2001-1}{Living Rev.\ Rel.\  {\bf 4}, 1 (2001)}
  [\href{\arxiv/astro-ph/0004075}{astro-ph/0004075}]; \,\ 
  S.~Tsujikawa,
  \href{\doi/10.1007/978-90-481-8685-3_8}{Dark Matter and Dark Energy. Astrophysics and Space Science Library, vol 370. Springer, Dordrecht},
  \href{\arxiv/arXiv:1004.1493}{arXiv:1004.1493} [astro-ph.CO].

\bibitem{Comelli:2012db}
  D.~Comelli, M.~Crisostomi and L.~Pilo,
  \href{\doi/10.1007/JHEP06(2012)085}{JHEP {\bf 1206} (2012) 085}
  [\href{\arxiv/arXiv:1202.1986}{arXiv:1202.1986} [hep-th]].

\bibitem{DeFelice:2017oym}
  A.~De Felice, S.~Mukohyama and J.~P.~Uzan,
  \href{\doi/10.1007/s10714-018-2342-z}{Gen.\ Rel.\ Grav.\  {\bf 50} (2018) no.2,  21}
  [\href{\arxiv/arXiv:1702.04490}{arXiv:1702.04490} [hep-th]].
  
  
\bibitem{deRham:2010eu}
  C.~de Rham and A.~J.~Tolley,
  \href{\doi/10.1088/1475-7516/2010/05/015}{JCAP {\bf 1005} (2010) 015}
  [\href{\arxiv/arXiv:1003.5917}{arXiv:1003.5917} [hep-th]].
  
\bibitem{Horndeski:1974wa} 
  G.~W.~Horndeski,
  \href{\doi/10.1007/BF01807638}{Int.\ J.\ Theor.\ Phys.\  {\bf 10}, 363 (1974)}.

\bibitem{Nicolis:2008in} 
  A.~Nicolis, R.~Rattazzi and E.~Trincherini,
  \href{\doi/10.1103/PhysRevD.79.064036}{Phys.\ Rev.\ D {\bf 79}, 064036 (2009)}
  [\href{\arxiv/arXiv:0811.2197}{arXiv:0811.2197} [hep-th]].
  
\bibitem{Deffayet:2009wt} 
  C.~Deffayet, G.~Esposito-Farese and A.~Vikman,
  \href{\doi/10.1103/PhysRevD.79.084003}{Phys.\ Rev.\ D {\bf 79}, 084003 (2009)}
  [\href{arxiv/arXiv:0901.1314}{arXiv:0901.1314} [hep-th]].
  

  
\bibitem{Amendola:2006we}
  L.~Amendola, R.~Gannouji, D.~Polarski and S.~Tsujikawa,
  \href{\doi/10.1103/PhysRevD.75.083504}{Phys.\ Rev.\ D {\bf 75} (2007) 083504}
  [\href{\arxiv/gr-qc/0612180}{gr-qc/0612180}].

\bibitem{Maldacena:2002vr}
  J.~M.~Maldacena,
  \href{\doi/10.1088/1126-6708/2003/05/013}{JHEP {\bf 0305} (2003) 013}
  [\href{\arxiv/astro-ph/0210603}{astro-ph/0210603}].
      
\bibitem{Kobayashi:2011nu}
  T.~Kobayashi, M.~Yamaguchi and J.~Yokoyama,
  \href{\doi/10.1143/PTP.126.511}{Prog.\ Theor.\ Phys.\  {\bf 126} (2011) 511}
  [\href{\arxiv/arXiv:1105.5723}{arXiv:1105.5723} [hep-th]].

\bibitem{TheLIGOScientific:2017qsa}
  B.~P.~Abbott {\it et al.} [LIGO Scientific and Virgo Collaborations],
  \href{\doi/10.1103/PhysRevLett.119.161101}{Phys.\ Rev.\ Lett.\  {\bf 119} (2017) no.16,  161101}
  [\href{\arxiv/arXiv:1710.05832}{arXiv:1710.05832} [gr-qc]].
  
\bibitem{Monitor:2017mdv}
  B.~P.~Abbott {\it et al.} [LIGO Scientific and Virgo and Fermi-GBM and INTEGRAL Collaborations],
  \href{\doi/10.3847/2041-8213/aa920c}{Astrophys.\ J.\  {\bf 848} (2017) no.2,  L13}
  [\href{\arxiv/arXiv:1710.05834}{arXiv:1710.05834} [astro-ph.HE]].

\end{thebibliography}
\end{document}